# Nanocomposite membranes with Au nanoparticles for dialysis-based catalytic reduction-separation of nitroaromatic compounds

Piotr Cyganowski*, Joanna Wolska

Department of Process Engineering and Technology of Polymer and Carbon Materials

Faculty of Chemistry, Wroclaw University of Science and Technology

Wyb. S. Wyspianskiego 27, 50-370 Wrocław

Correspondence to: P. Cyganowski, e-mail: piotr.cyganowski@pwr.edu.pl, phone: 71 320 23 83

ORCID: 0000-0002-3110-4246

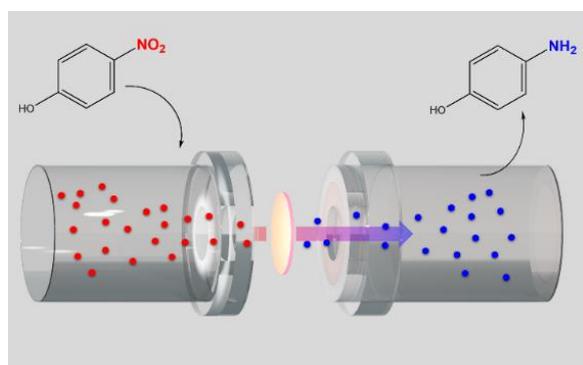

**Abstract**

Apart from the fact that nitroaromatic compounds (NARs) have toxic and mutagenic characteristics, they are also essential substrates for the synthesis of aromatic amines (AAMs). In this context, the present study presents a new approach that enables NAR-contaminated wastewaters to be treated as a reagent for the synthesis of AAMs. It involves the fabrication of anion exchange membranes with Au nanoparticles (AuNPs) that simultaneously reduce 4-nitrophenol (4-NP) and separate the resultant 4-aminophenol (4-AP) via. the dialysis mechanism. The nanocomposite membranes were prepared by amino-modification of poly(vinyl chloride) films obtained in the presence of cyclohexanone (CH) or tetrahydrofuran (THF), followed by Au(III) reduction coupled-adsorption. The nanomaterials were analysed using scanning transmission electron microscopy (STEM) and Fourier-transformation infrared spectroscopy (FT-IR). The catalytic reaction was carried out in a dialysis unit, where the concentration of 4-NP in the wastewater, and the concentration of separated 4-AP were monitored using UV-Vis spectroscopy. The nanocomposite membranes formed using THF effectively reduced the 4-NP and separated the resultant 4-AP. The yield of the 4-NP conversion reached 80% with a rate constant of $11.30·10-3$ $min^{-1}$. Based on the results, THF contributed to the formation of diffusion paths in which the 4-NP was simultaneously separated and reduced.

## 1. Introduction

Aromatic Amines (AAMs) are crucial for various branches of industry, and are key building blocks for the manufacturing of photographic films, corrosion retarders, polymers, herbicides, and dyes. They are also used in drug formulations, and are the key substrates for the production of large scale pharmaceuticals, including paracetamol, ibuprofen, acetaminophen, bicalutamide and nilutamide (antiandrogens), linezolid (an antibiotic for multidrug resistant Gram-positive bacteria), and fosamprenavir (the anti-HIV drug that was found to also be effective in the treatment of COVID-19) [1-4]. The most popular processes for the production of AAMs involve the reduction of nitro aromatic compounds (NARs) to AAMs *via*. non-catalysed or catalysed reactions [5-8]. However, NARs are able to covalently bind DNA, and they thus reveal carcinogenic characteristics and pose a risk of damaging the mechanisms of $O_2$ transfer in the human body [9]. Therefore, when considering NARs as fine substrates, which involves their common use with facilitated discharge into wastewaters, serious environmental and health hazards can be identified. However, when considering NAR-contaminated wastewaters as substrates for the production of AAMs, it is possible to meet these issues on two different levels. First, it could be possible to address major issues of NAR occurrence, and second, it could result in the production of AAMs in a process that would be environmentally sustainable. This challenge could be addressed using a nanotechnology-based approach.

The growing interest in the development of new nanomaterials (NMs) that reveal catalytic activity is mainly related to the increased efficiency and selectivity they show. Moreover, the unique optical, mechanical, chemical, thermal, and biological properties of NMs allow them to be applied in a number of scientific and industrial settings [10-16]. Among the many different NMs, nanocatalysts (NCats), which contain nanoparticles (NPs) of noble metals, are particularly desired. The reason for this is the well-documented efficiency of such NMs as AuNPs, PtNPs and PdNPs in the reduction of NARs [17-21]. Among these, AuNPs are particularly important, as they do not interact with olefinic substituents, and instead prefer the reduction of $–NO_2$ groups over other reducible moieties. This leads to their apparent selectivity. Moreover, AuNPs offer stability and the possibility of carrying out NAR reduction under mild conditions [20,22,23]. However, AuNPs do not offer separation of the resultant AAMs from the reaction media, and what is more, Au nanostructures, like any other nanomaterial, suffer from limited stability.

For all these reasons, we propose an entirely new approach within the present work, which will enable the simultaneous reduction of NARs and the continuous separation of the resultant AAMs. To achieve this, a set of nanocomposite membranes (nMs) with AuNPs were synthesized. The polymeric films, based on poly(vinyl chloride), were modified with amines to enable the *in-situ* synthesis of AuNPs *via.* the reduction-coupled adsorption of $AuCl_4^-$ ions on amino molecular reactors. In this scenario, the amino functionalities play the role of the reducing and capping agents for Au(III), and also enable dialysis-based separations to be performed. Thus, the AuNP-loaded nMs were used as nanocatalysts (NCats) for the reduction of 4-nitrophenol (4-NP) in a reduction-separation process. Within this, the 4-NP was reduced, while the resultant 4-aminophenol (4-AP) was separated *via.* a dialysis-based mechanism.

## 2. Experimental Procedure
### 2.1. Materials and methods

Poly(vinyl chloride) (PVC) powder (Ongrovil® S-5167 - a suspension polymer with a medium molecular weight and with a K of 66-68) was kindly supplied by BorsodChem. The initial PVC gel membranes were prepared by dissolving PVC in cyclohexanone (CH) or tetrahydrofuran (THF) supplied by Avantor Peformance Materials Ltd. (Gliwice, Poland)

The amines: 1,2-diaminoethane (DAE, >99%), and 2,6-diaminopiridine (DAP, 98%) were delivered by Sigma-Aldrich Chemical Co. (Poland branch), and used as received. The $HAuCl_4 \cdot 3H_2O$ used as the AuNPs' precursor was acquired from Sigma-Aldrich Chemical Co. (Poland branch) and dissolved in 0.1 mol $L^{-1}$ HCl in order to receive a stock solution with a concentration of Au(III) 1000 mg $L^{-1}$. A representative NAR, 4-nitrophenol (4-NP), was purchased in MERCK (Poland branch) and dissolved in RO water to obtain a 0.1 mmol $L^{-1}$ solution for the catalysis-dialysis process. All the other reagents were purchased in Avantor Peformance Materials Ltd. (Gliwice, Poland) and used as received.

The physiochemical characteristics of the PVC films, and the exchange membranes that are based on them, were performed by determining the Cl and N concentrations using Schrödinger's and Kjeldahl's procedures, respectively [24,25]. In the case of determining the Cl concentration, approximately 30 mg slice of a membrane was burned in $O_2$ atmosphere on a Pt catalyst. The resultant fumes were absorbed in 15 mL of 3% $H_2O_2$, that afterwards was titrated with the use of $NH_4SCN$ and $AgNO_3$ according to the Vohlard's protocol [24]. In turn, the N concentration was determined by mineralizing approx. 0.1 g slice of a membrane in concentrated $H_2SO_4$. Then, the so-prepared solution was distilled with water steam intro methyl red–bromocresol green indicator and titrated using HCl. The presence of amino functionalities introduced into the PVC was confirmed using Attenuated Total Reflectance Fourier Transformation Infrared Spectroscopy (FT-IR) by analysing the spectra recorded with the aid of the JASCO ATR-FTIR 4700 (MD, USA) instrument. The concentration of Au was assessed by mineralization of the AuNPs-loaded membranes in concentrated $HNO_3$, followed by the analysis of Au concentration with the aid of Flame Atomic Absorption Spectroscopy (FAAS) using the GBC Avanta spectrometer (VIC, Australia). The morphology of the AuNPs loaded into the nMs was assessed by Transmission Electron Microscopy (TEM) by capturing photomicrographs with the aid of the FEI Tecnai $G^2$ 20 X-Twin instrument (ThermoFisher Scientific) equipped with the Energy-Dispersive X-ray Spectrometer (EDX) and Selected Area Electron Diffractometer (SAED). The analysis was performed using Transmission (TEM) or Scanning-Transmittion (STEM) modes. The static contact angles of the water were measured using a goniometer PG-X (Fibro Systems). The catalytic activity of the nMs was assessed using the JASCO V-570 (MD, USA) UV-Vis spectrophotometer by recording spectra in the range of 200-600 nm. The dialysis cell used for catalysis-dialysis process was produced using Formlabs Form3 3D printer (MA, USA), and the Clear4 resin (Formlabs).

### 2.2. Synthesis of the nanocomposite membranes

First, two PVC gel films with a thickness of 0.25 mm were prepared. The membranes were formed using the inversion phase process (the dry variant of the method).The PVC powder was dissolved in THF or CH to obtain two solutions, within which PVC constituted 8 wt.%. Then, the PVC films were cast at room temperature from tetrahydrofuran solution (PVC-THF) or cyclohexanone solution (PVC-CH) onto glass plates using casting knife with a 0.25 mm slit. After this, the casted solutions were left for 72 h at ambient temperature to allow the THF and CH to evaporate. The precipitated PVC (dry phase separation) on a glass plate was then immersed in water and gently separated from the glass base. .

The so-prepared PVC-CH and PVC-THF films were then modified using DAE and DAP. Based on our previous works, which focused on heterogenous catalysts [26-28], it can be stated that the greatest efficiency of Au(III) reduction is achieved when no additional crosslinking due to the $-NH_2$ functionalities is observed. Therefore, the modification was carried out under mild conditions in the following way. The PVC-CH and PVC-THF films were cut in order to obtain samples with approximate dimensions of 5x5 cm. Then, the so-prepared scraps were swollen overnight in 25 mL of methanol, after which DAE or DAP were added. The volume of amines was adjusted in a way that allowed a 6:1 amine–to–Cl (in PVC) molar ratio to be obtained. Afterwards, the so-prepared mixtures were placed on an orbital shaker and left for 14 days, after which the amino-modified PVC membranes were washed and introduced with water, and then subsequently with 1 mol $L^{-1}$ HCl and NaOH to finally wash it with water.

The resultant PVC-CH-DAE, PVC-THF-DAE, PVC-CH-DAP, and PVC-THF-DAP samples were then used in their swollen form for the actual synthesis of the nMs.

The loading of the AuNPs was performed using an *in-situ* approach previously developed for suspension copolymers [26-28]. The process involves the reduction-coupled adsorption of $AuCl_4^-$ anions on amino functionalities. Then, due to the electron transfer from N atoms, and the autooxidation of the water [28], the Au(III) reduces and precipitates onto amino functionalities. Therefore, the amino-modified membranes were introduced into 25 mL of 1000 mg Au $L^{-1}$ in 0.1 mol $L^{-1}$ HCl solution in order to carry out the anion exchange of $AuCl_4^-$. Then, the Au-saturated membranes were washed and introduced into 25 mL of RO water, where the Au(III) reduced and precipitated directly in the polymeric network, after which the AuNPs-loaded membranes were used for further studies The simplified procedure is displayed in Figure 1.

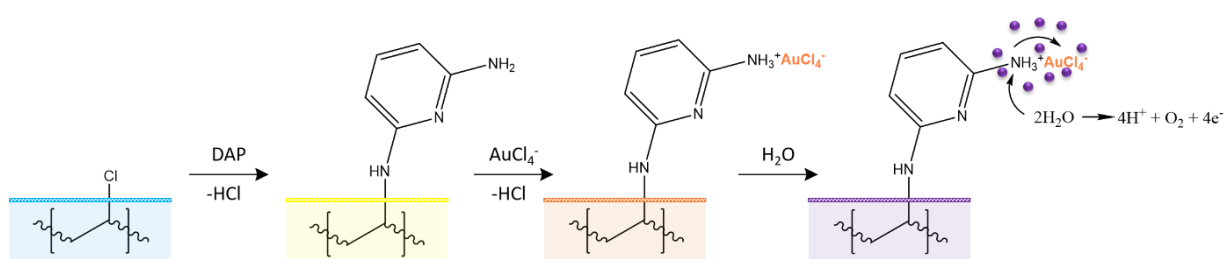

**Figure 1.** Simplified scheme for the modification of the PVC films and the loading of the AuNPs in the amino-modified membranes (example for DAP)

### 2.3. Research on catalytic activity

The obtained nMs PVC-CH-DAE@Au, PVC-THF-DAE@Au, PVC-CH-DAP@Au, and PVC-THF-DAP@Au were used for the catalysis-dialysis process for the reduction of the 4-NP with simultaneous separation of the resultant 4-AP.

The process was carried out in a dialysis cell designed for this purpose, which is displayed in the Graphical Abstract and Figure 4. The device consists of two dialysis chambers, which are separated by a nM with AuNPs and which have an assured contact surface of 4.9 $cm^2$. The procedure was carried out in the following way. To the left chamber, 35 mL of the feed solution containing 0.1 mmol $L^{-1}$ of 4-NP was introduced. Then, 2.5 mL of 0.1 mol $L^{-1}$ $NaBH_4$ was added and mixed with the feed solution. Meanwhile, the right chamber was filled with 35 mL of RO water to act as the stripping solution. The process was carried out for 1h while applying constant mixing with a magnetic rod rotating horizontally.

The actual assessment of the catalysis-dialysis process was performed using UV-Vis spectroscopy by recording the corresponding spectra in both the feed and stripping solutions. The species, including the 4-NP, 4-nitrophenolate anion and 4-NA were detected at the $\lambda_{max}$ 318, 400 and 295 nm, respectively. Then, a pseudo-first order kinetic model was applied to calculate the rate constant ($k_1$). This was done by drawing plots $lnA_t/A_0$ versus $t$ (where $A$ is absorbance at 400 nm and $t$ is time). The $k_1$ was taken from the slope of the above-mentioned plots. Simultaneously, the stripping solution was monitored by observing a band at 295 nm to verify whether the resultant 4-AP had passed through the nM.

## 3. Results and discussion
### 3.1. Synthesis of the anion exchange membranes

The anion exchange membranes were based on the PVC films, and were formed with the aid of CH or THF as a solvent. Then, after evaporation of the solvent, the PVCs were modified into anion exchange membranes using DAE and DAP. The initial assessment of the synthesis and modification was carried out by determining the concentrations of the Cl and N in the samples. The corresponding values are given in Table 1.

**Table 1.** Characteristics of the membranes

| Polymeric base | Solvent | Amine | Cl[a] | Cl[b] | N[c] | Au[d] |
|---|---|---|---|---|---|---|
| PVC | CH | - | 22.2±1.5 | - | - | - |
|  |  | DAE | 19.5±0.5 | 2.8±0.6 | 3.6±1.7 | 10.1±1.1 |
|  |  | DAP | 21.0±0.9 | 0.5±0.4 | 1.2±0.9 | 8.3±0.7 |
|  | THF | - | 22.7±2.0 | - | - | - |
|  |  | DAE | 2.5±1.1 | 22.0±2.1 | 57.2±1.9 | 21.3±2.1 |
|  |  | DAP | 17.5±0.2 | 5.9±0.9 | 20.2±1.2 | 18.3±1.2 |

[a]concentration of covalently bonded Cl (mmol g$^{-1}$); [b]concentration ionic Cl; [c]concentration of N (mmol g$^{-1}$); [d]amount of Au loaded into a membrane (%)
PVC: poly(vinyl chloride); CH: cyclohexanone; THF: tetrahydrofuran; DAE: 1,2-diaminoethane; DAP:2,6-diaminopyridine

Based on the results, it can be observed that the efficiency of the modification (based on covalently bonded Cl, Table 1) was greater in the case of the PVC films formed in the presence of THF than when formed in the presence of CH (89% *vs.* 10% Cl substitution). This effect is attributed to the fact that THF is recognized as a better solvent for PVC when compared to CH [29][30-32]. Thus, PVC-THF seems to be more accessible to modification when compared to PVC-CH. This conclusion is supported by the concentration of N (Table 1). Generally, the DAE and DAP particles migrated easier in the PVC-THF matrix, leading to the greater concentration of functional groups. In contrast, the modification of the PVC-CH films was not so efficient, as the Cl substitution in the case of the PVC-CH-DAE was negligible. It was also observed that the modification was greater in the case of the DAE when compared to the DAP (based on Cl and N concentrations, Table 1). This phenomenon is attributed to the size and shape of the amino molecules. The aliphatic DAE diffused easier in the PVC when compared to the aromatic, amino-substituted DAP. As a result, the obtained anion exchange membranes revealed a differentiated concentration of functionalities. Based on the N concentration (Table 1), it can be stated that almost no modification was observed in the case of the PVC-CH-DAP, while the most efficient modification was observed in the case of the PVC-THF-DAE membrane.

Because of the significant differences in modification yields, the presence of functionalities derived from the DAE and DAP was confirmed using ATR FT-IR spectroscopy. The spectra of the PVC films, as well as their amino-derivatives, are displayed in Figure 2.

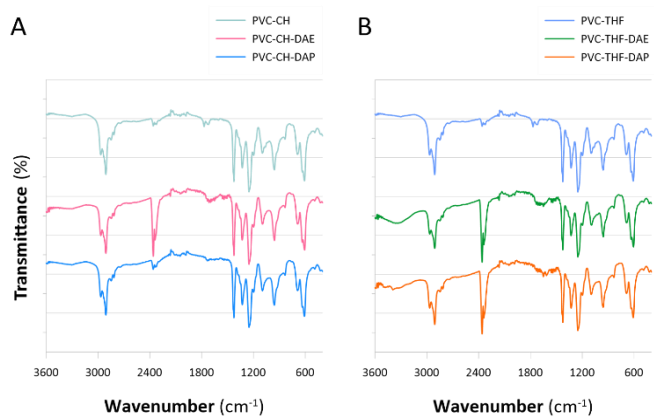

**Figure 2.** ATR-FTIR spectra of membranes derived from the PVC films using (**A**) cyclohexanone (CH) and (**B**) tetrahydrofuran (THF)

The spectra of the unmodified PVC (PVC-CH, PVC-THF) displayed bands that are characteristic for aliphatic CH shifted by $CH_2$ deformations to 1425 and 1423 cm$^{-1}$. In addition, the bands attributed to the CH shifted by C–Cl vibrations can be observed at 586 (PVC-CH) and 585 cm$^{-1}$ (PVC-THF) [33]. The spectra recorded for the modified samples are consistent with the data displayed in Table 1. The spectra of the PVC-CH-DAE and PVC-CH-DAP samples did not reveal bands at ~3300 cm$^{-1}$ that are characteristic for N–H stretching in primary and secondary amines, which is reflected by the relatively small N concentration (Table 1). However, the spectrum of the PVC-CH-DAE displayed a very strong band at 2059 cm$^{-1}$, which was assigned to the presence of amino hydro-halides, *i.e.* protonated −NH groups [33]. This confirms the presence of dissociated DAE particles in the PVC-CH. In turn, no similar effect was observed in the case of the DAP modification. This, linked with the data collected in Table 2, suggests that the DAP-modification of the PVC-CH was probably unsuccessful. A completely different situation was observed in the spectra of the PVC-THF-DAE and PVC-THF-DAP. They clearly revealed wide bands at 3045 and 3051 cm$^{-1}$, which is attributed to the N–H stretching, respectively. What is more, both of these samples revealed bands at 2036 and 2040 cm$^{-1}$, which is assigned to the amino hydro-halides [33]. This confirms the primary role of the THF in the further modification of the PVC samples.

Because the synthesized membranes are designed for the catalytic process based on the dialysis separation, yet another important factor is H$_2$O contact angle (measure of wettability) thereof. For this reason, the PVC films before and after modification were assessed in the context of H$_2$O droplets contact angle they revealed. The corresponding values were: 75° for PVC-THF, 70° for PVC-CH, and ~60° for the remaining samples, i.e. PVC-THF-DAE, PVC-THF-DAP, PVC-CH-DAE, and PVC-CH-DAP. Based on these results, it could be stated that the greatest difference in the measured H$_2$O contact angle was in the case of PVC-THF (75° vs. 60°). This is yet another support for the conclusion on the preferable role of THF in the synthesis of anion exchange membrane.

### 3.2. Synthesis and loading of the AuNPs

All of the obtained membranes (irrespective of the extent of modification) were employed in the reduction-coupled adsorption of AuCl$_4^-$. As a result of the process (according to the mechanism displayed in Figure 2), the Au(III) ions were reduced and precipitated directly into the PVC polymeric matrix. As can be seen in Table 1, the amount of Au loaded into the anion exchange membranes ranged from 8.3±0.7 to 21.3±2.1%, and was proportional to the concentration of amines present in the polymeric films. This phenomenon was expected, as it has been already noticed, that the amines play primary role in the reduction and capping of AuCl$_4^-$ [26,27]. Consequently, despite much poorer yield of modification using DAP, the difference in Au concentration (irrespectively from the membrane) was not so great as compared to the polymers modified with DAE (Table 1). This was attributed to the aromatic character of the functionality derived from DAE, that has been already identified to increase efficiency of Au(III) reduction [26,27]. The morphology of the resultant nMs: Au@PVC-CH-DAE, Au@PVC-CH-DAP, Au@PVC-THF-DAE, Au@PVC-THF-DAP, as well as the AuNPs they contain, were investigated using TEM and STEM. The corresponding photomicrographs are displayed in Figure 3.

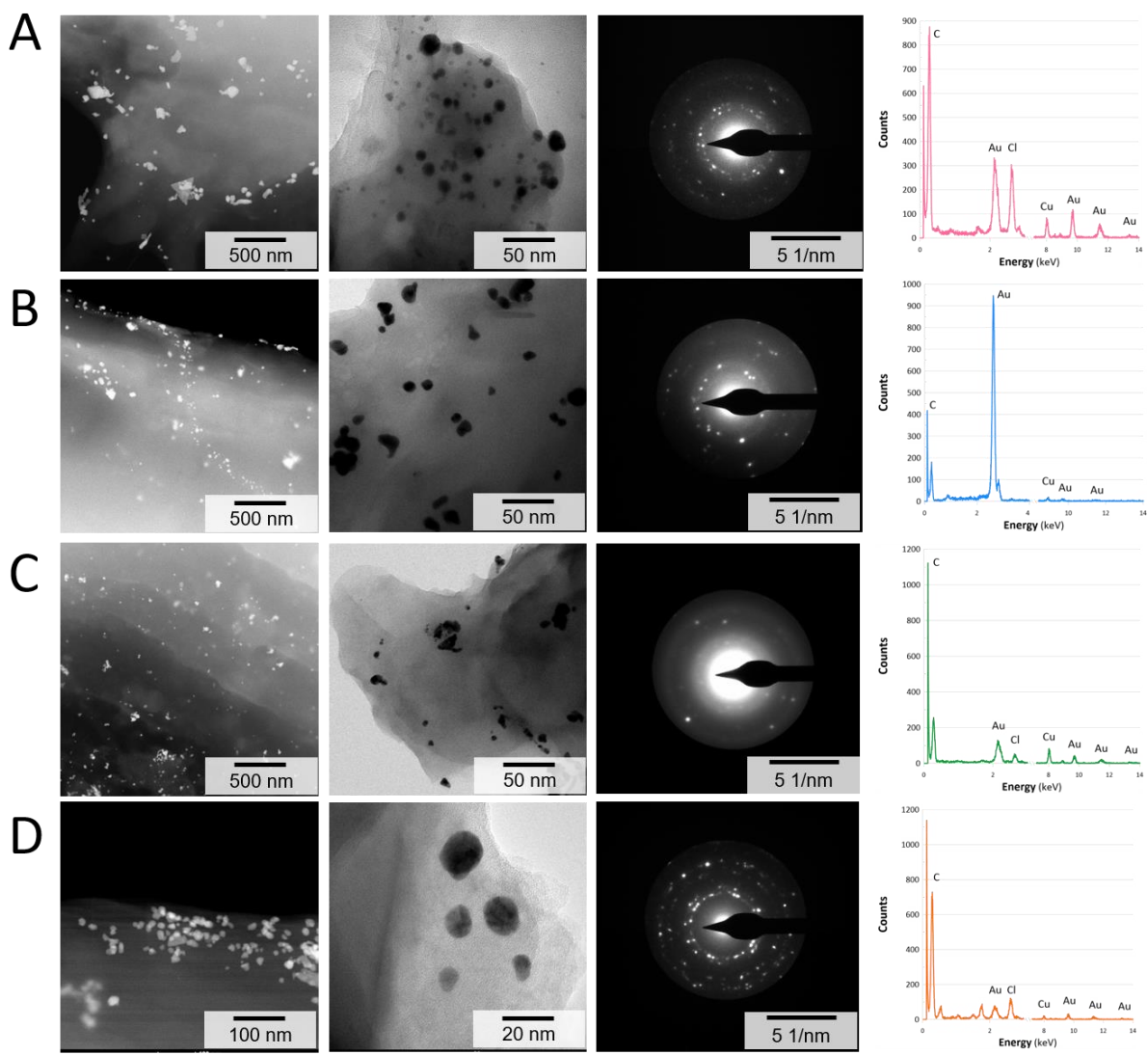

**Figure 3**. Photomicrographs (in order): STEM, TEM, SAED, and EDX of (**A**) Au@PVC-CH-DAE, (**B**) Au@PVC-CH-DAP, (**C**) Au@PVC-THF-DAE, and (**D**) Au@PVC-THF-DAP

The analysis confirms that the nanostructures loaded into the nMs are indeed AuNPs. Based on the SAED spectra, the calculated d-spacings for all the samples are approx. 1.24, 1.50, 2.12 and 2.52 Å, which is consistent with previous literature reports focusing on AuNPs of different shapes [34]. Based on the STEM and TEM photomicrographs, it can be stated that the AuNPs are well-dispersed, with a size ranging from 2 to 50 nm. Moreover, larger structures, with the size range of 50-200 nm, can be observed near the surface of the polymer. This effect was expected, and was caused by the concentration gradient defined by the difference of the number of saturated functionalities in and out the polymer's structure [26]. At the same time, no significant difference between the AuNPs obtained using DAE and DAP was observed. The lack of difference between the AuNPs was achieved deliberately by ensuring the presence of $-NH_2$ groups in both polymers. Therefore, in the present studies, the differentiated solvents and amines used for the formation and modification of the PVC films will be considered as a vital factor for the dialysis-separation processes.

### 3.3. The catalysis-dialysis process

The obtained nMs with AuNPs were used in the unique process, within which the continuous reduction of the 4-NP and the separation of the 4-AP was carried out. Such an approach allows environmentally harmful 4-NP to be reduced, and a fine chemical product – 4-AP – to be produced. Within this scenario, simulated wastewaters containing 4-NP were contacted with anion exchange nMs Au@PVC-CH-DAE, Au@PVC-CH-DAP, Au@PVC-THF-DAE, and Au@PVC-THF-DAP. These nanomaterials play a universal and multifunctional role, and are catalysts and dialysis membranes at the same time. The concept of the catalysis-dialysis process is displayed in Figure 4.

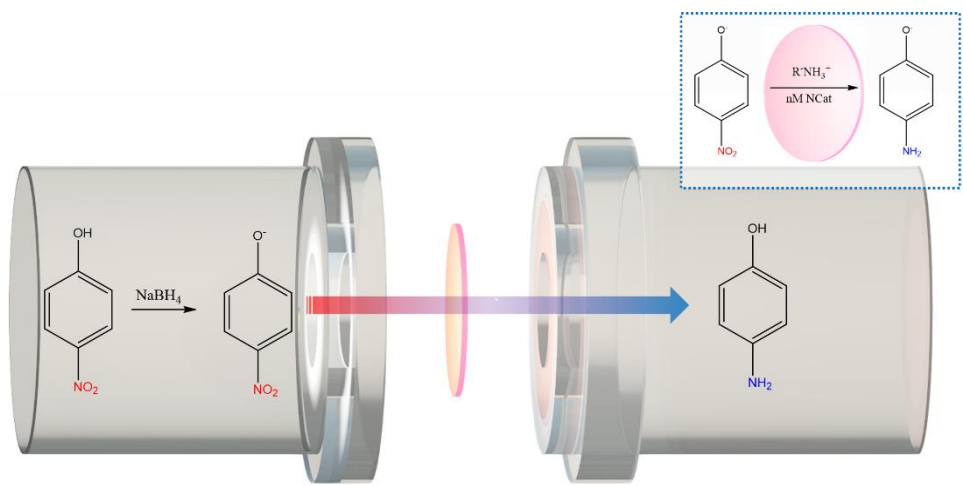

**Figure 4.** Concept of the unique catalysis-dialysis process

First, 4-NP was transformed into the 4-nitrophenolate anion, which is attracted by a nM due to the presence of amino functionalities within its structure. Then, the 4-nitrophenolate anion was transferred through the polymer to the stripping solution. However, in this case, the dialysis membrane is a nanocomposite containing AuNPs. Therefore, 4-nitrophenolate anion was reduced to 4-AP upon its contact and passage through the membrane. Simultaneously, the content of the feed solution and the stripping solution was monitored using UV-Vis spectroscopy. The recorded spectra for both dialysis chambers are displayed in Figure 5.

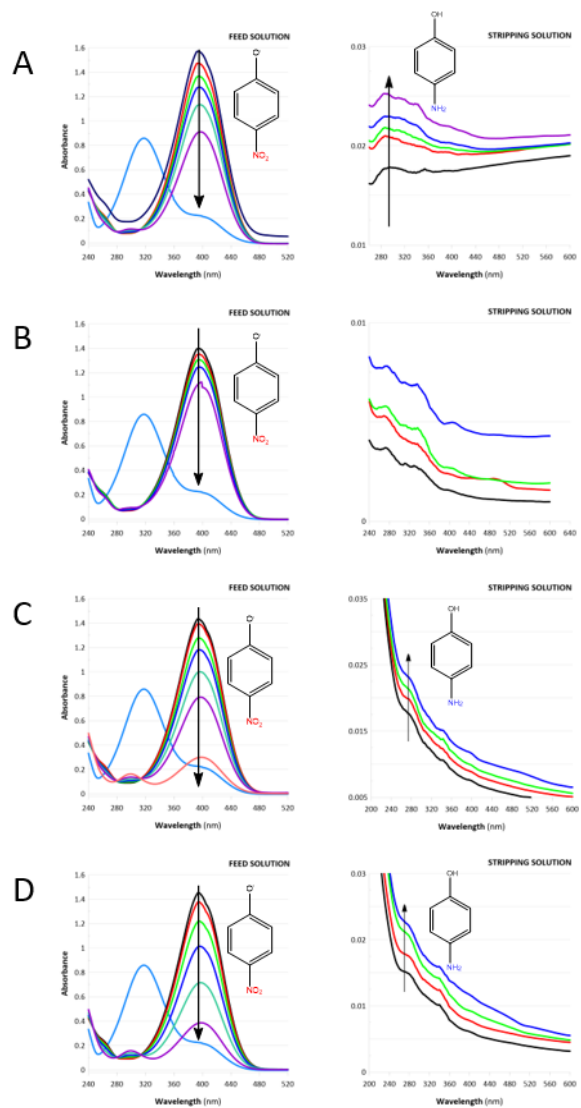

**Figure 5.** UV-Vis spectra of the feed and stripping solution upon contact with (**A**) Au@PVC-CH-DAE, (**B**) Au@PVC-CH-DAP, (**C**) Au@PVC-THF-DAE, and (**D**) Au@PVC-THF-DAP. (**A**) Reaction time (RT): 240 min; 4-NP conversion: 43%. (**B**) RT: 170 min, 18%. (**C**) RT: 212 min, 80%. (**D**) RT: 120 min, 73%

Based on the decreasing absorbance at $\lambda_{max}$ 400 nm, each of the nMs catalysed the reduction of the 4-NP. However, significant differences between the efficiency of the process were observed. The membranes obtained using CH as a solvent (Au@PVC-CH-DAE and Au@PVC-CH-DAP, Figure 5A-B) converted 18-43% of the 4-NP within 240 and 170 min, respectively. At the same time, the membranes formed in the presence of THF (Au@PVC-THF-DAE and Au@PVC-THF-DAP, Figure 5C-D) converted 80 and 73% of the 4-NP within 212 and 120 min, respectively. The calculated pseudo-first order kinetic constants were (in order of decrease): $11.30 \cdot 10^{-3}$ min$^{-1}$ for the Au@PVC-THF-DAP, $5.58 \cdot 10^{-3}$ min$^{-1}$ for the Au@PVC-THF-DAE, $2.25 \cdot 10^{-3}$ min$^{-1}$ for the Au@PVC-CH-DAE, and $1.51 \cdot 10^{-3}$ min$^{-1}$ for the Au@PVC-CH-DAP. Based on these values, it can be clearly stated that the membranes formed using CH reveal a much smaller catalytic activity when compared to the nMs obtained with the aid of THF. This phenomenon is probably attributed to the number of functionalities introduced into the PVC. As described above, the efficiencies of modification in the case of the materials formed using CH was poor (see Table 1 and Figure 2 for details). This had an impact on the ability of the membranes to transfer anions through their structure, and therefore the greater amount of amino ligands in the Au@PVC-THF-DAE and Au@PVC-THF-DAP easily attracted 4-NP to their surface, where it was reduced. Conversely, the Au@PVC-CH-DAE and Au@PVC-CH-DAP samples, although containing AuNPs, probably generated a much smaller driving force (electrostatic interactions) and were unable to attract 4-NP as well as the samples obtained with the aid of THF. This conclusion is supported by the UV-Vis spectra of the stripping solutions (Figure 5) obtained using Au@PVC-CH-DAP. In this case, although the reaction was catalysed, the dialysis effect was not achieved (Figure 5B). Due to the fact that this sample contained AuNPs (Figure 3B), the phenomenon can be linked with the lack of amino functionalities, which in turn stopped the ability of the membrane to transfer anions. In turn, the Au@PVC-CH-DAE obtained with the aliphatic DAE revealed a greater concentration of N (3.6 mmol g$^{-1}$), and thus enabled separation of the 4-AP (Figure 5A). This is evidenced by the bands at $\lambda_{max}$ 280-295 nm [28]. To finally exclude the influence of other factors, the nature of the surfaces of the nMs was also verified by measuring the $H_2O$ contact angle for each of the samples. In all cases, the contact angle was ~60°, indicating that no significant differences in the surface properties were observed. This allows a conclusion that the observed differences must be solely linked with the ability of the membrane to transfer anions.

At the same time, the process carried out using the Au@PVC-THF-DAE and Au@PVC-THF-DAP seems to be more efficient in terms of both catalysis and dialysis. This draws yet more support for the increased efficiency of PVC membranes obtained with the aid of THF. As was expected, the THF, as a better solvent towards PVC when compared to CH, probably formed "diffusion paths". These in turn enabled efficient amino modification, that was not limited only to the surface of a membrane, but to its inner part as well. This led to the formation of molecular catalysis-dialysis channels within which the attracted 4-nitrophenolate is reduced and transferred to the stripping solution as the reaction product, *i.e.* 4-AP.

The difference between the dialysis efficiency achieved by the obtained nMs allows to predict the mechanism of the catalysis-dialysis process. Based on the UV-Vis spectra displayed in Figure 5 it can be noted that the resultant product of 4-NP reduction, *i.e.* 4-AP appears in the feed solution (Figure 5, $\lambda_{max}$ 280-295 nm). This suggests that the actual catalytic reduction occurs before actual dialysis-driven separation. After this, with observable delay, the reaction product appears in the stripping solution only if actual dialysis occur (see Figure 5, right column). Based on these observations, the possible mechanism of the novel catalysis-dialysis process could look as follows. First, the 4-NP in form of $O_2N-Ar-OH$ is contacted with $NaBH_4$ used as the reducing agent. This forces the transformation of 4-NP to 4-nitrophenolate anion, that occurs in the form of $O_2N-Ar-O^-$ [35,36]. Second, the 4-nitrophenolate anion is attracted to the surface of the membrane via. driving force of the process (concentration gradient), due to the electrostatic interactions at solid-liquid interface. In this case, the solid membrane is loaded with AuNPs forcing the reduction of 4-nitrophenolate anion under mild conditions [20]. This resulted in the formation of 4-AP in the feed solution (as evidenced by UV-Vis spectra, Figure 5). Third, due to the excess of $NaBH_4$ the 4-AP occurs in the form of $H_2N-Ar-O^-$, that is able to be transferred through a nM *via.* dialysis mechanism (anion-exchange-driven concentration gradient) do the stripping solution

At this point, it is also worth to mention the stability of the AuNPs loaded into the membranes. In this context, the solutions remaining after all of the catalytic tests were checked for their potential ability to scatter light as well as the presence of a concentration of Au (attributed to the leached AuNPs). Both of these tests were negative, which allowed to conclude that the AuNPs did not migrated out of polymeric network. Additionally, the membranes used for the process were also mineralized and checked in the context of Au concentration. In this case, no significant differences were observed.

## 4. Conclusion

Within the present study, we successfully obtained a series of nanocomposite membranes (nMs) with AuNPs. These materials efficiently reduced 4-NP and simultaneously separated the resultant 4-AP *via.* the dialysis mechanism. The Au@PVC-THF-DAP membrane revealed the greatest 4-NP reduction rate constant $k_1=11.30 \cdot 10^{-3}$ min$^{-1}$. This effect was attributed to the application of the THF solvent and its primary role in the formation of PVC films. It was concluded that the THF probably led to the conformation of polymeric chains, and thus led to the formation of diffusion channels that enables efficient modification and dialysis-based separation.

The literature provides only a few examples of nMs for the reduction of 4-NP. This includes egg-shell membranes with AgNPs that revealed $k_1=0.25$ min$^{-1}$ [37]. More complex approaches involving the application of a β-lactoglobulinfibrils-based membrane with a Cu-Ag-Au NPs alloy offer 4-NP conversion with $k_1=17.1$ min$^{-1}$. However, in these cases, the concept did not involve the separation of 4-AP. Alternatively, there are also examples of nMs that offer a separation functionality. For instance, porous PAN-Si-Cu-Ag enabled the separation of 4-NP with its simultaneous reduction ($k_1$ unknown) [38]. In turn, the porous PVDF membrane with AuNPs [39] offered up to 90% of 4-NP conversion with $k_1=10^3$ min$^{-1}$. An alternative approach involves the application of a cellulose-based fibrous membrane with AgNPs [40], or a nanofibrous PAN membrane with AuNPs [41], which leads to the complete reduction of 4-NP with $k_1=0.195 \cdot 10^3$ min$^{-1}$ and $k_1=0.028$ min$^{-1}$, respectively. However, the membranes mentioned above enable separation of the resultant 4-AP *via.* the filtration mechanism. Therefore, successful 4-NP reduction and 4-AP separation strictly depends on the process parameters. However, there is a considerable risk of 4-AP contamination. Moreover, when considering real wastewaters containing various contaminants besides 4-NP, the so-prepared materials may not be selective enough, even though they can enable 4-NP and 4-AP separation.

The nMs developed in the present study address the above-mentioned issues. They enable catalytic separation *via.* the dialysis mechanism. Therefore, neither non-ionic nor cationic species can pass through the membrane. Because of this, the presented solution provides a very significant advantage when compared to other studies. The developed nMs offer 4-NP reduction and molecular-level separation of the obtained 4-AP, and thus may be a very tempting alternative to the solutions already reported.

The catalysis-dialysis concept links the major challenges arising from the occurrence of NARs in the environment and the synthesis of essential AAMs. As such, the developed nMs with AuNPs enable 4-NP contaminated wastewaters to be treated as a reagent for the production of 4-AP. Because the mechanism of the catalysis-dialysis process is universal, it could be easily developed to fabricate a new set of nanomaterials that reveal anion exchange properties and which contain NPs of different metals. This may result in the development of new methods, in turn making the achievement of environmental sustainability easier.

## 5. Acknowledgement

This work was supported by the National Science Centre (Poland) within the project UMO-2020/39/D/ST8/01352 granted to Piotr Cyganowski. This work was also supported by the subsidy granted to the Wroclaw University of Science and Technology by Polish Ministry of Education and Science. Piotr Cyganowski is also supported by Polish Ministry of Education and Science under the programme for outstanding young scientists.

## 6. Author contributions

**Piotr Cyganowski:** Conceptualization, Data curation, Formal analysis, Funding acquisition; Investigation, Methodology, Project administration, Resources, Supervision, Validation, Visualization, Writing - original draft, Writing - review & editing; **Joanna Wolska:** Methodology, Writing - review & editing

## 7. Data availability

The raw/processed data required to reproduce these findings cannot be shared at this time as the data also forms part of an ongoing study.

## 8. References

[1] H.K. Kadam, S.G. Tilve, Advancement in methodologies for reduction of nitroarenes, RSC advances 5(101) (2015) 83391-83407.

[2] B. Vellingiri, K. Jayaramayya, M. Iyer, A. Narayanasamy, V. Govindasamy, B. Giridharan, S. Ganesan, A. Venugopal, D. Venkatesan, H. Ganesan, COVID-19: A promising cure for the global panic, Sci. Total Environ. (2020) 138277.

[3] A. Provenzani, P. Polidori, Covid-19 and drug therapy, what we learned, Int. J. Clin. Pharm. (2020) 1.

[4] S. Nandi, P. Patel, H.K. Noor-ul, A.V. Biradar, R.I. Kureshy, Nitrogen-rich graphitic-carbon stabilized cobalt nanoparticles for chemoselective hydrogenation of nitroarenes at milder conditions, Inorg. Chem. Front. 5(4) (2018) 806-813.

[5] H. Wei, X. Liu, A. Wang, L. Zhang, B. Qiao, X. Yang, Y. Huang, S. Miao, J. Liu, T. Zhang, FeO x-supported platinum single-atom and pseudo-single-atom catalysts for chemoselective hydrogenation of functionalized nitroarenes, Nature communications 5 (2014) 5634.

[6] A. Corma, P. Serna, Chemoselective hydrogenation of nitro compounds with supported gold catalysts, Science 313(5785) (2006) 332-334.

[7] H.-U. Blaser, A. Indolese, A. Schnyder, Applied homogeneous catalysis by organometallic complexes, Curr. Sci. (2000) 1336-1344.

[8] R.A. Sheldon, H. Van Bekkum, Fine chemicals through heterogeneous catalysis, John Wiley & Sons2008.

[9] G.F. Smith, Designing drugs to avoid toxicity, Prog. Med. Chem., Elsevier2011, pp. 1-47.

[10] R.C. Zhang, D. Sun, R. Zhang, W.-F. Lin, M. Macias-Montero, J. Patel, S. Askari, C. McDonald, D. Mariotti, P. Maguire, Gold nanoparticle-polymer nanocomposites synthesized

by room temperature atmospheric pressure plasma and their potential for fuel cell electrocatalytic application, Sci. Rep. 7 (2017) 46682.

[11] P. Cyganowski, A. Dzimitrowicz, P. Jamroz, D. Jermakowicz-Bartkowiak, P. Pohl, Polymerization-Driven Immobilization of dc-APGD Synthesized Gold Nanoparticles into a Quaternary Ammonium-Based Hydrogel Resulting in a Polymeric Nanocomposite with Heat-Transfer Applications, Polymers 10(4) (2018) 377.

[12] S. Sarkar, E. Guibal, F. Quignard, A.K. SenGupta, Polymer-supported metals and metal oxide nanoparticles: synthesis, characterization, and applications, J. Nanopart. Res. 14(2) (2012) 715.

[13] M. Berger, Nano-society. Pushing the boundaries of technology RCS Publishing, Cambridge, 2009.

[14] J.J. Ramsden, Nanotechnology: an introduction, Elsevier, New York, 2011.

[15] P. Baglioni, E. Carretti, D. Chelazzi, Nanomaterials in art conservation, Nature Nanotechnology 10 (2015) 287.

[16] M.J. Pitkethly, Nanomaterials – the driving force, Mater. Today 7(12, Supplement) (2004) 20-29.

[17] Y.-C. Chang, D.-H. Chen, Catalytic reduction of 4-nitrophenol by magnetically recoverable Au nanocatalyst, J. Hazard. Mater. 165(1-3) (2009) 664-669.

[18] Q. Wang, W. Jia, B. Liu, A. Dong, X. Gong, C. Li, P. Jing, Y. Li, G. Xu, J. Zhang, Hierarchical structure based on Pd (Au) nanoparticles grafted onto magnetite cores and double layered shells: enhanced activity for catalytic applications, Journal of Materials Chemistry A 1(41) (2013) 12732-12741.

[19] S.K. Ghosh, M. Mandal, S. Kundu, S. Nath, T. Pal, Bimetallic Pt–Ni nanoparticles can catalyze reduction of aromatic nitro compounds by sodium borohydride in aqueous solution, Applied Catalysis A: General 268(1-2) (2004) 61-66.

[20] P. Zhao, X. Feng, D. Huang, G. Yang, D. Astruc, Basic concepts and recent advances in nitrophenol reduction by gold-and other transition metal nanoparticles, Coord. Chem. Rev. 287 (2015) 114-136.

[21] A. Dzimitrowicz, P. Cyganowski, P. Pohl, W. Milkowska, D. Jermakowicz-Bartkowiak, P. Jamroz, Plant Extracts Activated by Cold Atmospheric Pressure Plasmas as Suitable Tools for Synthesis of Gold Nanostructures with Catalytic Uses, Nanomaterials 10(6) (2020) 1088.

[22] A. Abad, P. Concepción, A. Corma, H. García, A collaborative effect between gold and a support induces the selective oxidation of alcohols, Angew. Chem. Int. Ed. 44(26) (2005) 4066-4069.

[23] Y.T. Woo, D.Y. Lai, Aromatic amino and nitro–amino compounds and their halogenated derivatives, Patty's Toxicology (2001) 1-96.

[24] A.I. Vogel, Handbook of Quantitative Inorganic Analysis, Longman, London, 1978.

[25] C. Kjeldahl, A new method for the determination of nitrogen in organic matter, Z. Anal. Chem. 22 (1883) 366.

[26] P. Cyganowski, Fully recyclable gold-based nanocomposite catalysts with enhanced reusability for catalytic hydrogenation of p-nitrophenol, Colloid. Surface. A accepted manuscript (DOI: 10.1016/j.colsurfa.2020.125995) (2021).

[27] P. Cyganowski, D. Jermakowicz-Bartkowiak, A. Leśniewicz, P. Pohl, A. Dzimitrowicz, Highly efficient and convenient nanocomposite catalysts produced using in-situ approach for decomposition of 4-nitrophenol Colloid. Surface. A 590 (2020) 124452.

[28] P. Cyganowski, A. Lesniewicz, A. Dzimitrowicz, J. Wolska, P. Pohl, D. Jermakowicz-Bartkowiak, Molecular reactors for synthesis of polymeric nanocomposites with noble metal nanoparticles for catalytic decomposition of 4-nitrophenol, J. Colloid Interface Sci. 541 (2019) 226-233.

[29] G. Grause, S. Hirahashi, H. Toyoda, T. Kameda, T. Yoshioka, Solubility parameters for determining optimal solvents for separating PVC from PVC-coated PET fibers, J. Mater. Cycles Waste Manage. 19(2) (2017) 612-622.

[30] M. Macchione, J.C. Jansen, G. De Luca, E. Tocci, M. Longeri, E. Drioli, Experimental analysis and simulation of the gas transport in dense Hyflon® AD60X membranes: influence of residual solvent, Polymer 48(9) (2007) 2619-2635.

[31] L. Lapčík, V. Kellö, J. Očadlík, Kinetic study of dissolution of poly (vinyl chloride) in tetrahydrofuran cyclohexanone, cyclopentanone, and N, N-dimethylformamide, Chemical Papers 27(2) (1973) 239-248.

[32] J. Kostina, G. Bondarenko, M. Gringolts, A. Rodionov, O. Rusakova, A. Alentiev, A. Yakimanskii, Y. Bogdanova, V. Gerasimov, Influence of residual solvent on physical and chemical properties of amorphous glassy polymer films, Polym. Int. 62(11) (2013) 1566-1574.

[33] D.A. Long, Infrared and Raman characteristic group frequencies. Tables and charts John Wiley & Sons, Ltd., Chichester, 2004.

[34] A.M. Goswami, S. Ghosh, Biological synthesis of colloidal gold nanoprisms using Penicillium citrinum MTCC9999, Journal of Biomaterials and Nanobiotechnology 4(02) (2013) 20.

[35] R. Sedghi, M. M Heravi, S. Asadi, N. Nazari, M. R Nabid, Recently used nanocatalysts in reduction of nitroarenes, Curr. Org. Chem. 20(6) (2016) 696-734.

[36] K. Zhang, J.M. Suh, J.-W. Choi, H.W. Jang, M. Shokouhimehr, R.S. Varma, Recent advances in the nanocatalyst-assisted NaBH4 reduction of nitroaromatics in water, ACS omega 4(1) (2019) 483-495.

[37] M. Liang, R. Su, W. Qi, Y. Yu, L. Wang, Z. He, Synthesis of well-dispersed Ag nanoparticles on eggshell membrane for catalytic reduction of 4-nitrophenol, J. Mater. Sci. 49(4) (2014) 1639-1647.

[38] P. Li, Y. Wang, H. Huang, S. Ma, H. Yang, Z.-l. Xu, High efficient reduction of 4-nitrophenol and dye by filtration through Ag NPs coated PAN-Si catalytic membrane, Chemosphere 263 (2021) 127995.

[39] J. Wang, Z. Wu, T. Li, J. Ye, L. Shen, Z. She, F. Liu, Catalytic PVDF membrane for continuous reduction and separation of p-nitrophenol and methylene blue in emulsified oil solution, Chem. Eng. J. 334 (2018) 579-586.

[40] T.K. Das, S. Remanan, S. Ghosh, N.C. Das, An environment friendly free-standing cellulose membrane derived for catalytic reduction of 4-nitrophenol: A sustainable approach, J. Environ. Chem. Eng. 9(1) (2021) 104596.

[41] Y. Liu, G. Jiang, L. Li, H. Chen, Q. Huang, T. Jiang, X. Du, W. Chen, Preparation of Au/PAN nanofibrous membranes for catalytic reduction of 4-nitrophenol, J. Mater. Sci. 50(24) (2015) 8120-8127.